\documentclass[trackchanges,twocolumn]{aastex701}

\usepackage{hyperref}
\usepackage{float}

\begin{document}

\title{Early Emergence of Environmental Effects: Accelerated Galaxy Assembly in a $z=2.96$ Protocluster in the COSMOS Field}

\author[orcid=0009-0009-4837-2165]{Shuaiyi Li}
\affiliation{Department of Astronomy, Tsinghua University, Beijing 100084, China}
\email{lishuaiy25@mails.tsinghua.edu.cn}  

\author[orcid=0000-0001-8467-6478]{Zheng Cai} 
\affiliation{Department of Astronomy, Tsinghua University, Beijing 100084, China}
\email[show]{zcai@tsinghua.edu.cn}

\correspondingauthor{Zheng Cai}

\author[orcid=0000-0003-0111-8249]{Yunjing Wu} 
\affiliation{Kavli Institute for the Physics and Mathematics of the Universe (WPI), The University of Tokyo Institutes for Advanced Study, The University of Tokyo, Kashiwa, Chiba 277-8583, Japan}
\affiliation{Center for Data-Driven Discovery, Kavli IPMU (WPI), UTIAS, The University of Tokyo, Kashiwa, Chiba 277-8583, Japan}
\email{yunjing.wu@ipmu.jp}

\author[orcid=0000-0002-3489-6381]{Fujiang Yu} 
\affiliation{Department of Astronomy, Tsinghua University, Beijing 100084, China}
\email{yufj@mail.tsinghua.edu.cn}

\author[orcid=0000-0001-6052-4234]{Xiaojing Lin} 
\affiliation{Department of Astronomy, Tsinghua University, Beijing 100084, China}
\email{linxj21@mails.tsinghua.edu.cn}

\author[orcid=0009-0005-6104-6568]{Jingyang Men} 
\affiliation{Department of Astronomy, Tsinghua University, Beijing 100084, China}
\email{smenn0605@gmail.com}

\author[orcid=0009-0007-9647-9907]{Xiaoyang Wei} 
\affiliation{Department of Astronomy, Tsinghua University, Beijing 100084, China}
\email{xy-wei25@mails.tsinghua.edu.cn}

\begin{abstract}
The redshift range $z=2$--4 marks a critical transition in large scale structure formation, where the dynamically unrelaxed progenitors of local massive clusters undergo rapid stellar mass assembly. We report the discovery and physical characterization of a highly significant protocluster, PC J1001+0214, at z=2.96 within the COSMOS field. Leveraging the multi-wavelength COSMOS2025 catalog with exceptional photometric precision in conjunction with JWST/NIRCam wide-field slitless spectroscopy (WFSS) from the COSMOS-3D program, we robustly identify a cosmic overdensity with $\delta=2.66$. The structure comprises 131 member galaxies, including 21 spectroscopically confirmed members (identified primarily via He\,{\sc i}\,$\lambda10830$ emission) and 110 high-fidelity photometric members. A comparative analysis against a mass-complete coeval field sample reveals a statistically significant +0.2 dex shift in the stellar mass distribution of protocluster members, signaling accelerated mass assembly in the dense environment by $z\sim3$. While the protocluster population broadly follows the star-forming main sequence, low-to-intermediate mass galaxies ($\log_{10}(M_{*}/M_{\odot})\le9.7$) exhibit a measurable star formation rate (SFR) enhancement of +0.11 to +0.15 dex. Crucially, the quiescent fraction remains extremely low and indistinguishable from the field sample, implying that environmental quenching mechanisms have not yet become dominant. Furthermore, a preliminary size-mass analysis hints at elevated morphological compactness among protocluster members at fixed stellar mass. These results suggest that PC~J1001+0214 represents a growth-dominated protocluster phase in which environmental effects are already detectable, primarily through accelerated stellar mass assembly and ongoing growth rather than through strong quenching.
\end{abstract}


\section{Introduction} 
Galaxy properties in the local Universe are strongly dependent on environment, with massive, quiescent systems dominate the high-density cores of virialized clusters \citep{cluster_Dressler_1980,cluster_Kauffmann_2004,cluster_Percy_2003}. When and how such environmental dependencies manifest at higher redshift, especially at $z\sim2$--4 when cosmic star formation and stellar mass assembly are both highly active, remains a fundamental challenge in observational cosmology. \citep{Koyama_2013,Shimakawa_2015,importance_Sun_2024,Forrest_2024}
High-redshift protoclusters, the progenitors of present-day massive galaxy clusters, provide a unique temporal window into this problem. 
By observing these extreme overdensities at $z\ge2$, prior to cluster virialization and the onset of pervasive intra-cluster medium (ICM) shock-heating, protoclusters serve as premier laboratories to study the early effects of environment on star formation efficiency, metallicity, hierarchical mass assembly, and the onset of systemic quenching.
\citep{main_Overzier_2016, CC_property_2021, main_Wang_2022,importance_Sun_2024,main_Vine_2026}. 

Previous studies reveal a complex, often conflicting picture of environmental effects in high-redshift protoclusters. On one hand, dense environments can actively fuel early galaxy growth. Observations reveal enhanced star-formation activity in protoclusters relative to the coeval field \citep{main_Hayashi_2016, sfr_enhance_Shimakawa_2018,sfr_enhance_Shimakawa2_2018,sfr_enhance_Monson_2021,main_Wang_2022}. In contrast, some other overdense systems $z\sim2$--4 appear surprisingly mature. These structures already exhibit elevated quiescent fractions and early red-sequence formation \citep{parent_Shi_2021,intro_QG_Ito_2023,intro_QG_Tanaka_2024,main_Vine_2026}. This diversity implies that protoclusters span a broad range of evolutionary phases.  A larger sample of well-characterized systems is therefore required to systematically study different evolutionary stages and to understand how environmental effects emerge and operate in overdense regions \citep{importance_Chiang_2013,main_Overzier_2016,main_review}.

While protoclusters at high redshift are traditionally identified using biased tracers like radio galaxies, quasars, or narrowband emitters, these approaches are intrinsically selective. Deep photometric redshift surveys offer a more systematic and complete alternative for mapping overdensities across broad redshift ranges. However, given the uncertainties and catastrophic failures inherent to photometric redshift selections, as well as the coarse redshift resolution of photometric membership, spectroscopic validation still remains essential to confirm genuine physical structures \citep{main_Overzier_2016,intro_Cucciati_2018,Guaita_2020,Helton_2024}. Surveys that combine wide area, sufficient depth, accurate photometric redshifts, and efficient spectroscopic follow-up are required. COSMOS2025 and COSMOS-3D are particularly well suited to this need.
Based on James Webb Space Telescope (JWST) COSMOS-Web program (PID 1727) and ancillary multiwavelength data, COSMOS2025 delivers robust photometry, morphologies, and SED-based physical parameters for $780,000$ galaxies over the central 0.54 $\rm deg^2$ COSMOS field. With its exceptional photo-$z$ precision and improved mass completeness, COSMOS2025 provides a  more complete sample to search protoclusters \citep{cosmosweb_Casey_2023,cosmos25_Shuntov_2025}.  In addition to COSMOS2025, we also use COSMOS-3D survey for possible spectroscopic follow-up validation. COSMOS-3D is a JWST/NIRCam Wide Field Slitless Spectroscopy (WFSS) program  (PID 5893) within the COSMOS-Web field and is designed to provide slitless spectroscopic constraints on galaxies in the early Universe. It provides abundant 2D spectra for precise redshift verification, allowing us to confirm overdense structures.

In this work, we search for protocluster candidates in COSMOS2025 catalog and identify a significantly overdense structure, PC~J1001+0214. We use COSMOS-3D spectra to validate the structure and identify a clear redshift peak near $z\sim2.96$. We compare the protocluster members with a coeval field sample in terms of stellar mass distribution, the stellar formation rate (SFR)-stellar mass ($M_*$) relation, quiescent fraction, and the size-mass relation. This allows us to establish the reality of the structure, and to examine whether an overdense environment at $z\sim3$ already leaves observable signatures on galaxy growth and how these signatures manifest. This paper is organized as follows. In Section~\ref{sec2}, we describe the data and sample selection. In Section~\ref{sec3}, we present our searching method and main results. Finally, we summarize our results and discuss their implications in Section \ref{sec4}. All magnitude are expressed in the AB system \citep{AB_Oke_1974ApJS...27...21O}.

\section{Data and Reduction}\label{sec2}

\subsection{COSMOS2025 Photometric Catalog}\label{2.1}

Our analysis is mainly based on the COSMOS2025 catalog, which serves as the parent sample for the photometric identification of candidate protocluster galaxies.\footnote{The catalog is publicly available at \url{https://cosmos2025.iap.fr/}.} Built upon the JWST COSMOS-Web survey \citep{cosmosweb_Casey_2023}, COSMOS2025 combines deep near-infrared imaging from NIRCam in F115W, F150W, F277W, and F444W with parallel MIRI/F770W observations, together with extensive ancillary data from space- and ground-based facilities. The resulting catalog covers the central $0.54\,\mathrm{deg}^2$ of the COSMOS field and provides matched photometry over 37 bands spanning $0.3$--$8\,\mu$m, along with morphological measurements, photometric redshifts, and SED-based physical properties for nearly $780,000$ galaxies. 

The depth and wavelength coverage of COSMOS2025 are particularly suited to studies of galaxies at high redshift. The JWST observation reaches $5\sigma$ depths of 27.2--28.1 mag in the NIRCam bands, while the MIRI/F770W imaging extends over $\sim0.20\,\mathrm{deg}^2$ to 25.2 mag. Owing to this broad multiwavelength baseline and the improved near-infrared sensitivity, the catalog delivers highly reliable photometric redshifts. Comparison with spectroscopic compilations in \citet{compilation_Khostovan_2026} shows a precision of $\sigma_{\rm MAD}\approx0.012$ for sources with magnitude $m_{\rm F444W}<28$, remaining effective out to $z\sim9$ \citep{cosmos25_Shuntov_2025}. Here the median absolute deviation (MAD) is defined as:
\begin{equation}
    \sigma_{\rm MAD}=1.48\times \mathrm{median}\left(\frac{|\Delta z-\mathrm{median}(\Delta z)|}{1+z_{\rm spec}}\right),
\end{equation}
where $\Delta z = z_{\rm phot}-z_{\rm spec}$.

In addition, COSMOS2025 substantially improves the stellar-mass completeness relative to COSMOS2020, reaching an $\sim80\%$ completeness level at $\log_{10}(M_{\star}/M_{\odot})\sim9$ by $z\sim10$ and at $\log_{10}(M_{\star}/M_{\odot})\sim7$ near $z\sim0.2$ to a gain of roughly 1 dex over COSMOS2020 catalogs \citep{cosmos2020_Weaver_2022,cosmos2020_results_Weaver_2023,cosmos25_Shuntov_2025}. This combination of multi-band coverage, high-quality photo-$z$, and wide area makes COSMOS2025 an ideal foundation for systematically identifying and analyzing protocluster galaxies.

\subsection{Sample Selection}
We use the following selection criteria for COSMOS2025 to get a research sample with both completeness and accurate photometric redshift. We first restrict our analysis to galaxies located within the COSMOS-3D footprint expanded $5\arcmin$ round to enable subsequent validation with JWST/NIRCam WFSS spectroscopy. From this parent sample, we select galaxies with photometric redshifts $2\le z_{\mathrm{phot}}\le 4$ . At lower redshift, protoclusters are expected to be more evolved and approach the regime of mature clusters, whereas at higher redshift photometric redshift uncertainties often increase substantially, making overdensity identification and characterization less robust. 
To suppress contamination from catastrophic photometric redshift failures, we enforce a strict normalized uncertainty threshold of $\sigma_z/(1+z)\le 0.05$. 
Furthermore, we impose magnitude cuts of $\mathrm{m_{F150W}}\le 27.21$ or $\mathrm{m_{F444W}}\le 27.23$ to limit the rapid degradation of photo-$z$ precision at faint magnitudes \citep{criteria_mag_Darvish_2015}.
We then exclude all \texttt{STAR\_HSC} sources flagged in the HSC-based star mask, which identifies objects contaminated by bright stars in the HSC imaging, to ensure that the photometry and SED fitting results are reliable \citep{hscmask_Coupon_2017}. 

\begin{figure}
    \centering
    \includegraphics[width=\linewidth]{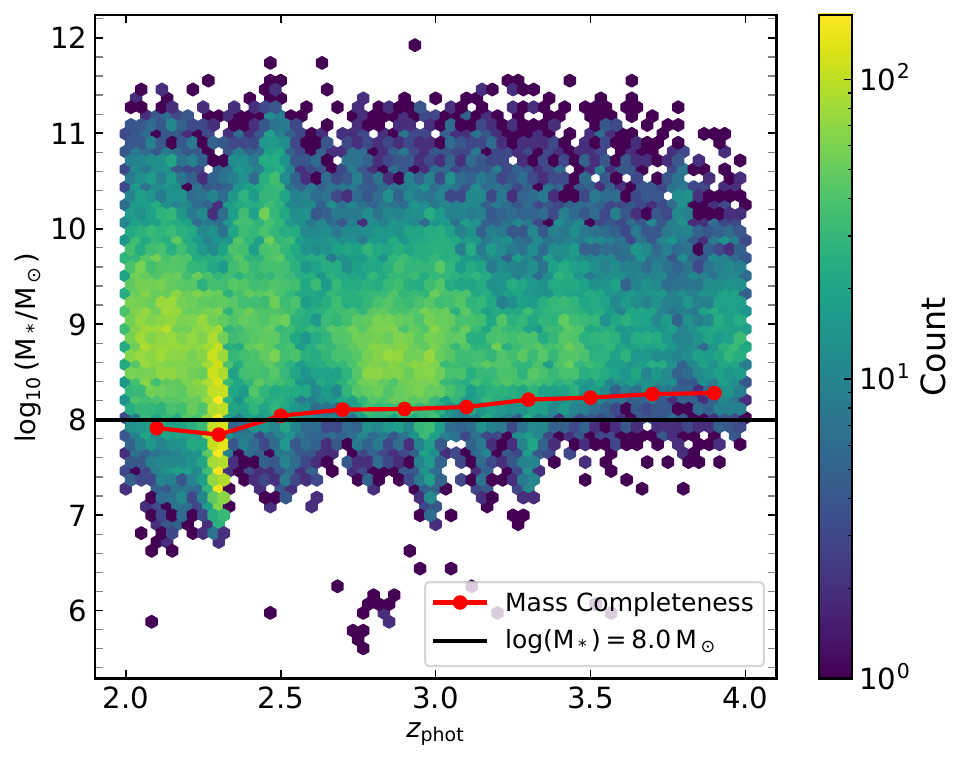}
    \caption{Stellar mass completeness in the $\log_{10}(M_\star/M_\odot)$ versus $z_{\rm phot}$ plane. The background color map shows the galaxy number counts. The red markers indicate the stellar mass completeness limits estimated in individual redshift bins following the method of \citet{mass_limit_Pozzetti_2010}. The black horizontal line marks the constant stellar mass threshold of $\log_{10}(M_\star/M_\odot)=8.0$ used in this work. }
    \label{fig:mass_completeness}
\end{figure}

To define a stellar-mass completeness limit, we follow the method of \citet{mass_limit_Pozzetti_2010} and \citet{cosmosweb_smf_Shuntov_2025}. In brief we select the 30\% faintest galaxies in different redshift bins, and rescale their stellar masses $M_*$ to an $\mathrm{F444W}$ magnitude limit of $m_{\rm lim}=27.5$ for COSMOS-Web survey using Eq. (\ref{mass}). Then we take the 90th percentile of the rescaled mass distribution as the stellar mass completeness limit. This yields the redshift dependent mass completeness shown as red markers in Figure~\ref{fig:mass_completeness}. Based on these estimates, we adopt $\log_{10}(M_\star/M_\odot)\ge 8.0$ as the stellar mass threshold throughout this work which can both balance completeness and a sufficiently large sample size for robust statistics.
\begin{equation}\label{mass}
    \mathrm{log_{10}}M_{m_{lim}}=\mathrm{log_{10}}M_{*}+0.4(m_{\rm F444W}-m_{\rm lim})
\end{equation}

\subsection{JWST/NIRCam Grism Spectroscopy}\label{sec2.3}
The COSMOS-3D program is a JWST Cycle-3 large program designed to conduct NIRCam WFSS observation with F444W filter and the grism R over $\sim0.33~ \rm deg^2$ within the COSMOS-Web footprint. In parallel with the long-wavelength (LW) WFSS exposures, the short-wavelength (SW) channel in F115W and F200W and LW channel in F356W obtain direct imaging , which is used to build the source catalog and anchor the spectral trace and wavelength mapping on the dispersed frames. 
With redcution of COSMOS-3D data, we can obtain 1D and 2D spectra of almost all galaxies in the effective field, enabling efficient spectroscopic confirmation for large galaxy samples. We reduce the grism exposures following the standard NIRCam/WFSS workflow described in \citet{grism_reduction_Sun_2023} and \citet{grism_reduction_Sun_2025}\footnote{The reduction code are availabel at \url{https://github.com/fengwusun/nircam\_grism}}. In brief, for each grism exposure and its corresponding SW direct image, we perform corrections including World Coordinate System (WCS) assginment, correcting for flat field, subtraction of a super sky background model and 1/f noise subtraction especially for spectra data.  We measure astrometric offsets between the SW direct images and external imaging catalogs, and then apply them to spectral tracing model. 2D spectra are stacked for each source, and 1D spectra are extracted using optimal extraction that accounts for the source morphology \citep{Horne_1986}.
       
We then use the semi-automated algorithm to determine the most possible redshift by identifying emission lines on continuum-subtracted grism spectra and also considering $z_{\rm phot}$. We refer to \citet{Lin_2026} for more details on the emission line searching algorithm.  We also perform visual inspection to reject residual contamination and artifacts. Then we select spectroscopically candidate members around the peak in grism redshift distribution.

\section{Results}\label{sec3}
\subsection{Surface Density Map and A Protocluster Candidate at z=2.96}
We search for overdense structures in the COSMOS-3D region using a grid-based fixed aperture counting method, and statistically correct for the loss of effective area caused by the \texttt{STAR\_HSC} mask \citep{pipeline_Toshikawa_2017}. We adopt overlapping sliding redshift windows with a step size of $\Delta z=0.1$ and a window width of 0.2, chosen to account for the typical photometric redshift uncertainties of our sample. We construct a regular grid with a spacing of $1\arcmin$ across the COSMOS-3D region. At each grid point, we define a circular counting aperture with a fixed angular radius of $r=5\arcmin$, which is motivated by \citet{radius_Lovell_2017} showing that the completeness and purity of protocluster galaxy populations are maximized ($>85\%$) at a radius of $10\pm2$~cMpc. Within each redshift slice, we count the raw number of galaxies inside each aperture and conduct mask region correction with the following methods, yielding a two-dimensional map of galaxy surface number density for that slice. 

Bright point sources and other artifacts contaminate substantial areas in the COSMOS field, leading to biased photometry and thus unreliable SED based properties. We therefore exclude galaxies in these contaminated regions when constructing our sample. However, without correction, such masking would systematically underestimate galaxy counts and introduce biases especially in the overdensity search. We then quantify the masked area fraction for each aperture using the mask region file from COSMOS2025 data release and a Monte Carlo area-estimation procedure, in which we uniformly sample random points within each counting aperture and compute the fraction falling inside the masked regions. Apertures with excessive masking can lead to unstable corrections, we thus discard apertures with masked area fraction $>50\%$. To obtain a robust estimate of the background counts and to correct for masking, we preferentially use reference apertures with minimal masked area fraction (defined here as $<15\%$). For these reference apertures, we estimate the mean number of galaxies in each slice dividing the raw counts by the unmasked area fraction  as the averaged number of total fields.
We then correct the raw count by adding a compensation term equal to the masked area fraction times the slice-averaged expected count for each aperture, which accounts for the galaxies statistically missing from the masked region and is subsequently used to obtain overdensity significance.
This approach avoids the noise amplification that would arise from directly dividing by unmasked area fraction for moderately masked apertures. Within each redshift slice, we compute the mean and standard deviation of corrected galaxy number, and define the standardized overdensity significance at each grid point as: 
\begin{equation}
\delta = \frac{N_{\mathrm{corr}}-\langle N_{\mathrm{corr}}\rangle}{\sigma(N_{\mathrm{corr}})}.
\end{equation}
where $N_{\rm corr}$ is corrected galaxy number within each aperture, $\langle N_{\mathrm{corr}}\rangle$ and  $\sigma(N_{\mathrm{corr}})$ is the corresponding mean and standard deviation of the whole COSMOS-3D field in each redshift slice \citep{delta_Toshikawa_2025}.
This yields a two-dimensional $\delta$ map for each redshift slice. We record the most significantly overdense regions i.e. the highest $\delta$ and subsequently validate them with JWST/NIRCam WFSS spectroscopy.

\begin{figure*}[t]
  \centering
  \begin{minipage}[t]{0.49\textwidth}
    \centering
    \includegraphics[width=\linewidth]{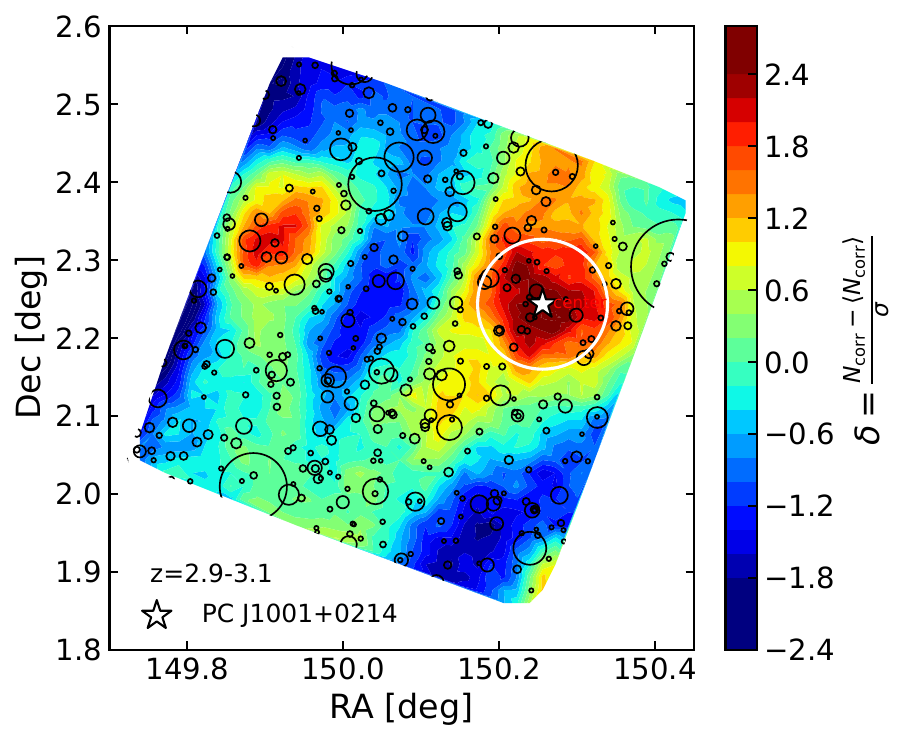}
  \end{minipage}\hfill
  \begin{minipage}[t]{0.49\textwidth}
    \centering
    \includegraphics[width=\linewidth]{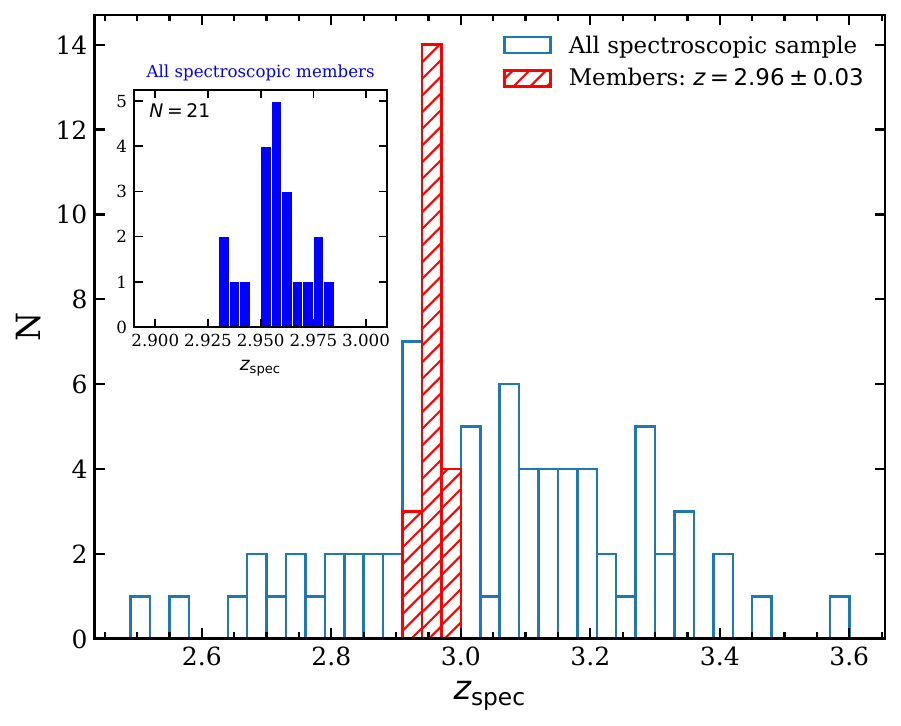}
  \end{minipage}

  \caption{\textit{Left}: Overdensity significance map in the $z_{\rm phot}=2.9$--$3.1$ slice. The color scale shows the $\delta=(N_{\rm corr}-\langle N_{\rm corr}\rangle)/\sigma$ . The black circles indicate the masked regions excluded from the analysis. The white star marks the center of the selected overdense region, PC~J1001+0214, at $(\mathrm{R.A.},\mathrm{decl.})=(150.2558^\circ,\,2.2432^\circ)$. This region reaches $\delta=2.66$ and is selected for subsequent spectroscopic validation and member galaxies analysis. \textit{Right}: Spectroscopic redshift distribution of galaxies in the selected overdense region. The open histogram shows the full 118 spectroscopic sample, while the hatched histogram marks galaxies within the adopted member range, $z=2.96\pm0.03$. The inset shows the redshift distribution of the spectroscopically confirmed members alone ($N=21$) with a clear peak near $z\sim2.96$.}
  \label{fig:overdense}
\end{figure*}

We rank the overdensity significance $\delta$ values from highest to lowest, and extract grism spectra for galaxies with methods described in Section \ref{sec2.3} within the corresponding apertures. In the $z_{\rm phot}=2.9-3.1$ slice, we identify an overdense region centered at $(\rm R.A.=150.2558^\circ,decl.=2.2432^\circ)$, whose $\delta=2.66$ lies in the top 0.51\% of all apertures in all slices. The left panel of  Figure~\ref{fig:overdense} shows the $\delta$ map in the $z_{\rm phot}=2.9-3.1$ slice. The white star and 5$\arcmin$ radius aperture marks this overdense region (hereafter PC~J1001+0214).  The map exhibits a clear local enhancement in $\delta$ around the center of  PC~J1001+0214, indicating that this region is not an isolated fluctuation but part of a coherent high density structure.

We choose this region to do further reaserach since galaxies in this region exhibit a high incidence of line emission features in their grism spectra, enabling spectroscopic redshift determination, validation and property analysis. 
To maximize the number of potential members, we extract grism spectra for all 840 galaxies with  $z_{\mathrm{phot}}  $ within $2.7<z<3.3$ in this overdense region and obtain grism spectroscopic redshifts of 118 galaxies. 
To validate our measurements, we cross-match our spectroscopic results against high-confidence ($\ge 85\%$) spectroscopic redshifts from the COSMOS compilation \citep{compilation_Khostovan_2026} \footnote{\url{https://github.com/cosmosastro/speczcompilation}} consisting of about 488k spectroscopic redshifts from past 20 years spectroscopic programs in COSMOS field and compilation of public NIRSpec spectra in the DAWN JWST Archive (DJA)\footnote{\url{https://dawn-cph.github.io/dja/index.html}}. Among the 13 matched galaxies, 12 yield an excellent agreement of $|\Delta z| \le 0.005$, decisively confirming the reliability of our grism $z_{\rm spec}$ derivation.
The right panel of Figure~\ref{fig:overdense} presents the spectroscopic redshift distribution of galaxies in the selected overdense region. It shows a prominent peak around $z\sim2.96$, supporting the presence of a real overdensity along the line of sight rather than a chance projected enhancement. 

\begin{figure*}[t]
  \centering
  \begin{minipage}[t]{0.49\textwidth}
    \centering
    \includegraphics[width=\linewidth]{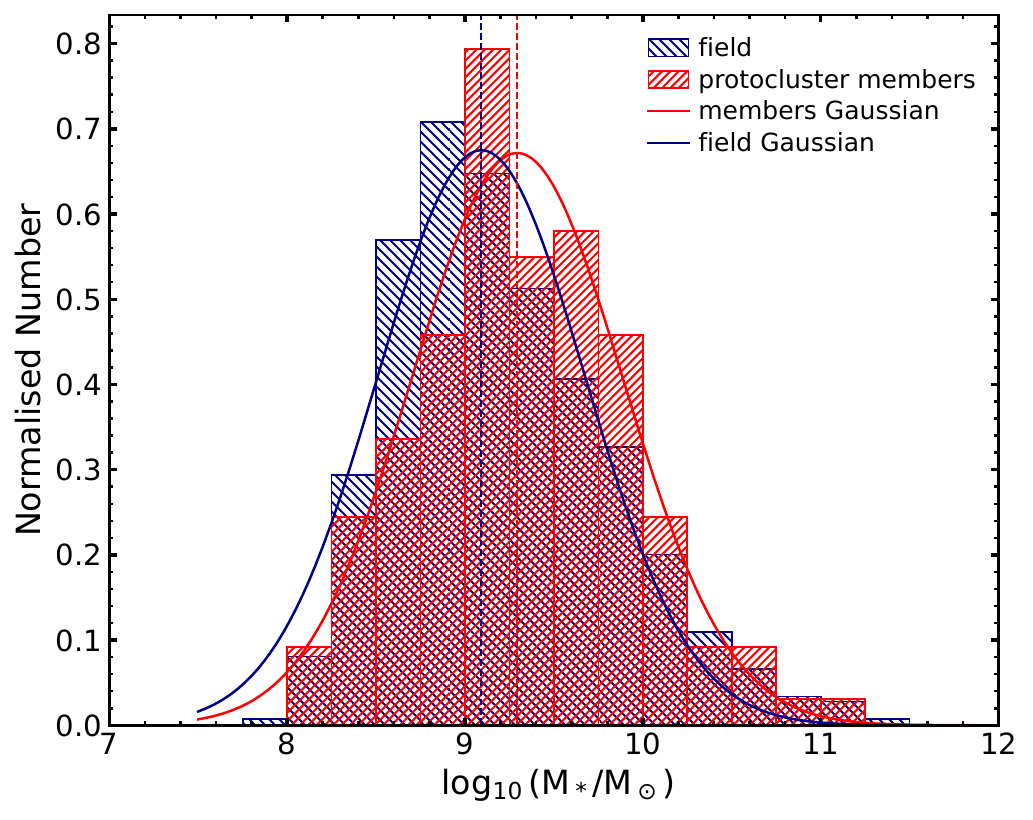}
  \end{minipage}\hfill
  \begin{minipage}[t]{0.49\textwidth}
    \centering
    \includegraphics[width=\linewidth]{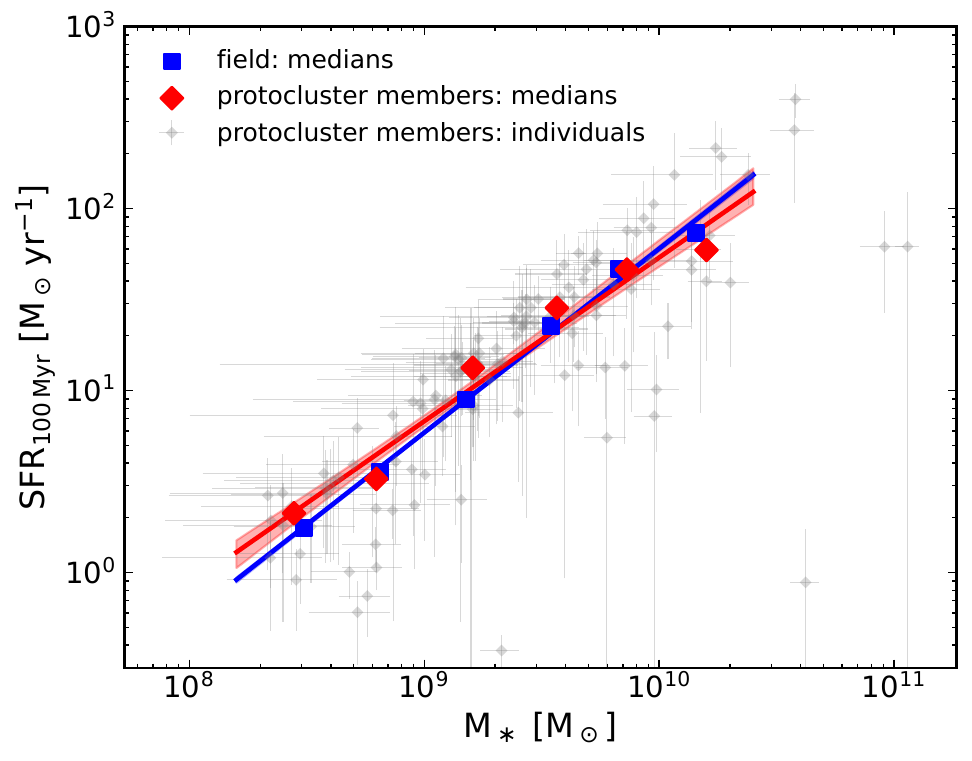}
  \end{minipage}
  \caption{\textit{Left}: Normalized stellar mass distributions of the protocluster members and field galaxies. The hatched red and blue histograms show the protocluster and field samples, respectively, while the  solid curves indicate the Gaussian fits. The vertical dashed lines correspond to peak masses of the fitted distributions. \textit{Right}: Stellar mass versus star formation rate averaged over the past 100 Myr derived from CIGALE. Gray diamonds represent individual protocluster members, with error bars from the $1\sigma$ photometric error. Red squares and blue diamonds mark the median SFRs of the protocluster and field samples in different stellar mass bins, respectively, while the corresponding curves show the best-fit relations for the two populations.
 The corresponding error bars are estimated by bootstrap resampling.}
  \label{fig:sfr}
\end{figure*}
\begin{deluxetable*}{ccccccccccc}
\tablecaption{Physical parameters of spectroscopic members\label{tab:spec_sample}}
\tabletypesize{\scriptsize}
\setlength{\tabcolsep}{3pt}
\tablehead{
\colhead{ID} &
\colhead{R.A.} &
\colhead{decl.} &
\colhead{$z_{\rm spec}$\tablenotemark{a}} &
\colhead{F115W} &
\colhead{F150W} &
\colhead{F277W} &
\colhead{F444W} &
\colhead{Emission Line\tablenotemark{b}} &
\colhead{$\log_{10}(M_\star/M_\odot)$} &
\colhead{SFR$_{100\,\mathrm{Myr}}$}
\\
\colhead{-} &
\colhead{(deg.)} &
\colhead{(deg.)} &
\colhead{-} &
\colhead{(AB mag)} &
\colhead{(AB mag)} &
\colhead{(AB mag)} &
\colhead{(AB mag)} &
\colhead{-} &
\colhead{-} &
\colhead{($M_\odot\,\mathrm{yr}^{-1}$)}
}
\startdata
665160 & 150.192961 & 2.198303 & \textbf{2.975} & 24.17 & 23.87 & 23.65 & 23.88 & HeI & $9.59 \pm 0.14$ & $49.13 \pm 23.25$ \\
675972 & 150.214127 & 2.261716 & 2.966 & 24.97 & 24.52 & 24.14 & 24.28 & HeI & $9.66 \pm 0.03$ & $56.93 \pm 13.68$ \\
679003 & 150.227951 & 2.279426 & 2.960 & 25.58 & 25.03 & 24.37 & 24.39 & HeI & $9.65 \pm 0.13$ & $13.76 \pm 7.38$ \\
683035 & 150.213834 & 2.311173 & 2.939 & 24.27 & 24.07 & 23.83 & 24.05 & HeI & $9.04 \pm 0.36$ & $8.94 \pm 7.72$ \\
683036 & 150.213869 & 2.311377 & 2.950 & 24.74 & 24.52 & 24.32 & 24.54 & HeI & $9.23 \pm 0.17$ & $19.38 \pm 12.14$ \\
708068 & 150.250940 & 2.172194 & 2.939 & 26.91 & 26.76 & 26.43 & 26.68 & HeI & $8.40 \pm 0.27$ & $1.82 \pm 1.29$ \\
711902 & 150.295468 & 2.181770 & 2.950 & 24.10 & 23.60 & 22.78 & 22.50 & HeI & $10.38 \pm 0.11$ & $153.12 \pm 49.08$ \\
713353 & 150.282920 & 2.197678 & 2.955 & 25.29 & 25.02 & 24.66 & 24.75 & HeI & $9.13 \pm 0.26$ & $15.83 \pm 13.04$ \\
713546 & 150.306917 & 2.190323 & 2.938 & 25.99 & 26.48 & 25.87 & 25.88 & HeI & $8.79 \pm 0.12$ & $1.43 \pm 0.35$ \\
713647 & 150.301877 & 2.192782 & 2.953 & 26.70 & 26.36 & 25.82 & 26.16 & HeI & $8.96 \pm 0.18$ & $2.35 \pm 1.31$ \\
717010 & 150.254623 & 2.232063 & 2.950 & 23.47 & 23.14 & 22.52 & 22.34 & HeI & $9.97 \pm 0.21$ & $51.61 \pm 22.23$ \\
717098 & 150.326510 & 2.206531 & 2.956 & 24.96 & 24.81 & 24.55 & 24.71 & HeI & $9.38 \pm 0.09$ & $24.03 \pm 6.08$ \\
720871 & 150.323823 & 2.233437 & 2.932 & 25.31 & 24.66 & 24.37 & 24.55 & HeI & $9.33 \pm 0.15$ & $14.53 \pm 7.82$ \\
723224 & 150.337441 & 2.245939 & 2.956 & 25.80 & 25.43 & 25.04 & 25.34 & HeI & $9.16 \pm 0.15$ & $15.13 \pm 7.87$ \\
724236 & 150.251235 & 2.284060 & \textbf{2.955} & 23.49 & 23.11 & 22.83 & 22.92 & HeI & $9.98 \pm 0.17$ & $105.79 \pm 65.29$ \\
724455 & 150.247561 & 2.286823 & 2.959 & 25.29 & 24.38 & 23.67 & 23.46 & HeI & $9.89 \pm 0.14$ & $42.83 \pm 26.59$ \\
725408 & 150.302892 & 2.272627 & 2.970 & 24.65 & 24.08 & 23.13 & 22.59 & HeI & $10.14 \pm 0.22$ & $51.45 \pm 29.74$ \\
725972 & 150.328446 & 2.266853 & 2.982 & 25.16 & 24.90 & 24.39 & 24.29 & HeI & $9.38 \pm 0.20$ & $25.18 \pm 14.95$ \\
726593 & 150.264357 & 2.293832 & \textbf{2.964} & 23.76 & 23.56 & 23.39 & 23.61 & HeI & $9.43 \pm 0.33$ & $25.59 \pm 23.58$ \\
727981 & 150.267049 & 2.302549 & \textbf{2.931} & 24.34 & 23.93 & 23.78 & 24.00 & HeI \& Pa$\delta$ & $9.44 \pm 0.17$ & $31.68 \pm 15.66$ \\
728264 & 150.294880 & 2.294512 & 2.961 & 24.09 & 23.42 & 22.68 & 22.45 & HeI & $10.27 \pm 0.14$ & $192.20 \pm 83.34$ \\
\enddata

\tablecomments{The IDs correspond to the \texttt{id} in the COSMOS2025 catalog. The R.A., decl., and four-band photometry are taken from COSMOS2025. The uncertainties on stellar mass and SFR are derived from the CIGALE SED fitting based on the $1\sigma$ photometric uncertainties.} 
\tablenotetext{a}{These redshifts are derived from the grism spectra. Values shown in bold denote sources cross-matched with high-confidence ($\geq 85\%$) spectroscopic redshifts from the COSMOS compilation \citep{compilation_Khostovan_2026}, with $|\Delta z| \leq 0.005$.}
\tablenotetext{b}{This column lists the emission lines detected in the grism spectra and used to determine $z_{\rm spec}$. HeI refers to He\,{\sc i}\,\(\lambda10830\), and Pa$\delta$ refers to the H\,I Paschen-$\delta$ line.}
\end{deluxetable*}

\subsection{Stellar Mass and Star Formation Rate}
To further investigate the galaxy population in PC~J1001+0214, we need to define a member sample for subsequent analysis. Spectroscopic members are selected as galaxies with $z_{\rm spec}$ within $z=2.96\pm0.03$  similar to the narrow $\Delta z\sim0.05$ membership ranges commonly adopted in spectroscopically confirmed protoclusters \citep{delta_Toshikawa_2025,Toshikawa2_2025}. We then conduct SED fitting using LEPHARE \citep{Lephare_Ilbert_2006} and CIGALE \citep{cigale_Boquien_2019} to further determine photometric members and analyse the environmental effects. 
We first tentatively identify 483 coarse photometric candidate members by filtering galaxies with the 16th and 84th percentile zphot within $z = 2.96\pm 0.25 $ just using the results of COSMOS2025 catalog. 
We then re-run the template-fitting code LePHARE for these candidates using a similar setup to COSMOS2025 plus restricting the redshift search to a narrow interval of $2.5<z<3.5$ around the targeted structure and using a finer redshift grid of $\Delta z=0.001$ to better sample the redshift probability distribution function. This reduces catastrophic solutions and provides more stable membership probabilities. The best-fit redshift $z_{\rm phot}$ is then adopted for photometric members selection. We determine photometric members as galaxies with their 16th and 84th percentile zphot newly obtained by LEPHARE within $z = 2.96\pm 0.15 $. We further use the rest-frame NUV$-r$ versus $r-J$ diagram derived from LEPHARE to exclude quiescent galaxies (QGs) \citep{QG_Ilbert_2013,QG_Edward_2023}. Finally we find 21 spectroscopically confirmed members with their information shown in Table~\ref{tab:spec_sample} and 110 photometric members in PC J1001+0214. For field galaxies, we re-conduct the same pipeline including re-running LEPHARE, selecting galaxies with the new 16th and 84th percentile zphot within $z = 3.0\pm 0.3 $, and excluding QGs. Then we get a field sample of 4185 field galaxies and 33 QGs in the stellar-mass range $M_{\star}>10^{10}\,M_{\odot}$, while we identify only 2 QGs among the 131 protocluster members, 
corresponding to quiescent fractions of $0.79^{+0.15}_{-0.13}\%$ and $1.53^{+1.55}_{-0.82}\%$, respectively, where the uncertainties denote $1\sigma$ binomial confidence intervals estimated using the Wilson score interval \citep{Wilson}.
For fitting physical parameters by CIGALE, we fix the redshift to the LEPHARE value for photometric members and to $z_{\rm spec}$ for spectroscopic members and performed Bayesian SED fitting using similar configurations to \cite{compilation_Khostovan_2026}. 

For LEPHARE SED-fitting, we fit the observed multi-band photometry with a representive library of galaxy SED templates based on the \citet{BC03} Stellar Synthetic Population models (hereafter BC03). We include three different attenuatin curves \citet{sed_Calzetti_2000,sed_Arnouts_2013,sed_Salim_2018}, set a maximum $\rm E(B-V)\le1$, allow for emission lines with a varable line flux factor of two, dust emission and intergalactic medium (IGM) absorption the same to \citet{cosmos25_Shuntov_2025}.  For CIGALE, we use a \citet{sed_IMF_Chabrier_2003} initial mass function (IMF) with BC03 stellar metallicities in the range of 0.0004 to 0.05 (0.02 – 2.5 $Z_{\odot}$). Then we assume a delayed-$\tau$ star formation history which includes a main stellar population and a young population formed in a recent burst. We define the main population age from 0.05 to 2.5 Gyr, e-folding timescale from 0.05 to 10 Gyr, age of the late burst from 5 to 50 Myr,  e-folding time of the late starburst population model from 5 to 100 Myr, and mass fraction of the late burst population between 0 and 90\%. We also include nebular emission line contribution. In brief, we set ionization parameter ${\rm log_{10}}U$ from -4 to -1, gas-phase metallicity from  0.0004 to 0.051 (0.02 – 2.5 $Z_{\odot}$), 100 $\rm cm^{-3}$ electron density and line width of  300 $\rm km\,s^{-1}$. We use a \citet{sed_Calzetti_2000} dust attenuation curve with no 2175 $\rm \AA$ bump and a maximum $\rm E (B - V) \le2$ and an LMC dust extinction curve \citet{sed_Pei_1992} for the emission lines. We caution that the physical parameters derived from SED fitting, including stellar mass and SFR, may be affected by systematic uncertainties related to the adopted model assumptions, such as the star formation history, dust attenuation law, and treatment of nebular emission. Nevertheless, because the protocluster and field samples are processed with the same photometry, fitting procedures, and model configurations, our analysis is primarily sensitive to relative differences between the samples, which should be more robust than the absolute calibration of the derived quantities.

Our results of stellar mass distribution and SFR-$M_\star$ relation are shown in Figure~ \ref{fig:sfr}. 
The left panel shows an analysis of the normalized stellar mass distributions revealing a distinct, top-heavy skew in the protocluster population relative to the coeval field. Parametrizing the distributions via Gaussian fitting yields a peak mass of $\log_{10}(M_\star/M_\odot)=9.294\pm0.035$ for protocluster members, contrasting with $9.092\pm0.034$ for field galaxies. This $\sim0.2$ dex positive offset signifies a fundamental divergence in the evolutionary timelines of the two environments, demonstrating that galaxies residing within the PC J1001+0214 overdensity have undergone highly accelerated, early hierarchical mass assembly.
The right panel presents a differential analysis of the star-forming main sequence revealing mass-dependent environmental modulation. Relative to the field main sequence, the protocluster population exhibits positive SFR offsets of +0.117 dex, +0.147 dex, and +0.115 dex across the low-to-intermediate stellar mass bins at $\log_{10}(M_\star/M_\odot)=8.2$–$8.6$, $9.0$–$9.4$, $9.4$–$9.7$ respectively with a negligible deviation ($-0.048$ dex) at the $8.6$–$9.0$ interval. Conversely, at the massive end, the offset diminishes to statistical insignificance because of the limited number of protocluster member galaxies. This mass-dependent divergence suggests that the overdense environment actively elevates the star formation efficiency of lower-mass systems potentially via enhanced cold gas accretion streams while more massive galaxies have already stabilized their star formation trajectories irrespective of their local density.

\subsection{Stellar mass-size relation}
\begin{figure}
    \centering
    \includegraphics[width=\linewidth]{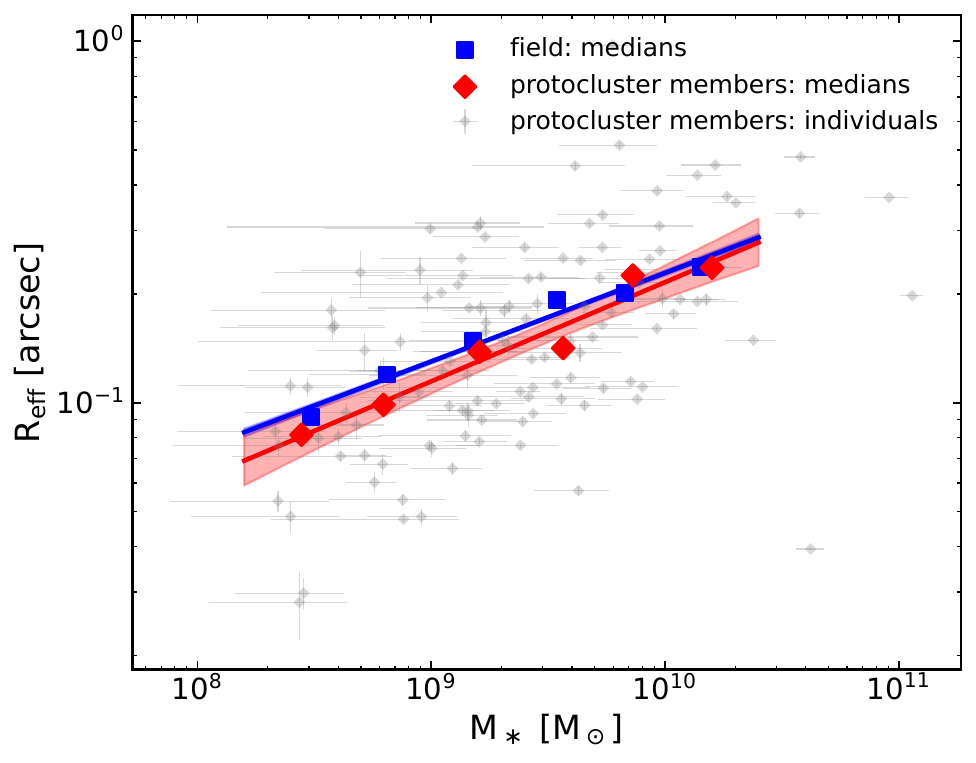}
    \caption{Effective radius versus stellar mass for the protocluster and field samples. Gray diamonds show individual protocluster members with mass errorbar derived from $1\sigma$ photometric error and radius errorbar derived from from the S\'ersic profile fitting in COSMOS2025 \citep{cosmos25_Shuntov_2025}. Red diamonds and blue squares mark the median $R_{\rm eff}$ of the protocluster and field galaxies in different stellar mass bins respectively, and the solid lines show the corresponding best-fit curve. The shaded bands indicate the uncertainties of bootstrap resampling. }
    \label{fig:size}
\end{figure}

As a tentative test of possible size differences between the protocluster and field populations, we compare their effective radii as a function of stellar mass. We directly adopt the COSMOS2025 catalog value \texttt{radius\_sersic} as the effective radius $R_{\rm eff}$. In COSMOS2025, source detection is based on  the four NIRCam bands, while the structural parameters of the single-S\'ersic models, including the effective radius, are obtained from simultaneous fitting to all NIRCam bands \citep{cosmos25_Shuntov_2025}.
Figure~\ref{fig:size} shows the resulting $R_{\rm eff}$-$M_\star$ relation for the protocluster members and the field sample.
Both samples exhibit the expected overall trend that galaxy size increases with stellar mass. The median relation of the protocluster members lies slightly below that of the field over most of the mass range, suggesting that galaxies in PC~J1001+0214 may be somewhat more compact. However, the current data provide only a tentative hint of smaller sizes among the protocluster members because of  the large scatter.
A more comprehensive assessment will require a larger sample and a more careful analysis that accounts for measurement uncertainties, and possible systematics in the catalog.

\section{Summary and Discussion}\label{sec4}
In this work, we report the discovery of a highly significant protocluster candidate, PC J1001+0214, at $z = 2.96$ identified using the COSMOS2025 multi-wavelength catalog and spectroscopically validated by JWST/NIRCam WFSS from the COSMOS-3D program. By robustly selecting member galaxies and comparing their physical properties with a coeval field sample, we investigate the environmental effects on galaxy evolution during the early stages of cluster assembly. Our main findings can be summarized as follows:
\begin{enumerate}
    \item The stellar mass distribution of the members is shifted toward higher masses, with the peak of the Gaussian fit higher by $\sim0.2$ dex than that of the field, indicating larger relative abundance of massive galaxies in the protocluster environment. 
    \item The SFR-$M_*$ relation shows that member galaxies broadly follow the same star-forming main sequence as field galaxies with only mild positive offsets at low and intermediate stellar masses $\log_{10}(M_\star/M_\odot)\le9.7$, while the difference becomes small at the massive end. 
    \item The fraction of quiescent galaxies at $M_{\star}>10^{10}\,M_{\odot}$ in the protocluster is extremely low and statistically indistinguishable from the field sample. 
    \item The size–mass relation shows a tentative tendency for protocluster galaxies to be smaller than field galaxies at fixed stellar mass maybe consistent with earlier or more centrally concentrated assembly, although the current scatter is too large for a firm conclusion.
\end{enumerate}
Taken together, these results suggest that PC~J1001+0214 is a protocluster at a stage when environmental effects are beginning to emerge, but have not yet fully transformed the galaxy population. The excess of massive galaxies indicates that stellar mass assembly may have proceeded earlier or more efficiently in the overdense environment than in the coeval field, consistent with accelerated galaxy growth in forming cluster environments. Meanwhile, the mild SFR enhancement, seen mainly at the low-to-intermediate mass end, suggests that the environmental influence is not a uniform elevation of star formation across the entire member population, but may instead preferentially affect galaxies that are still actively building up their stellar mass. This picture is consistent with the moderate enhancement reported in $z\sim2.5$ protoclusters by \citet{main_Hayashi_2016} and with the evidence for accelerated galaxy formation in the densest regions found by \citet{sfr_enhance_Shimakawa_2018}. At the same time, the very low quiescent fraction in both the protocluster and the coeval field implies that environmental quenching mechanisms associated with a mature hot intracluster medium, such as ram-pressure stripping or starvation, have likely not yet become dominant in this system \citep{results_Pan_2025,QG_Edward_2023}. The tentative compactness trend, if confirmed, would be broadly consistent with a more concentrated mode of assembly because of the enhanced interaction between galaxies in this protocluster.

Within the COSMOS field, PC~J1001+0214 can be placed into an evolutionary context by comparison with overdense structures at different redshifts. At earlier epochs, the Elent\'ari proto-supercluster at $z\sim3.3$ appears as a highly extended, still unvirialized structure composed of multiple overdense components distributed over tens of comoving Mpc, while Hyperion at $z\sim2.45$ represents a giant proto-supercluster on even larger scales \citep{intro_Cucciati_2018,Elent_Forrest_2023}. At more advanced stages, CL~J1001+0220 at $z=2.506$ already exhibits extended X-ray emission, indicating the early buildup of an intracluster medium \citep{CL_J1001+0220_Wang_2016}. In this context, PC~J1001+0214 at $z=2.96$ appears to trace a relatively early phase of a protocluster in which environmental effects are already becoming detectable. At the same time, its very low quiescent fraction suggests that the system has not yet reached the more advanced state such as CL~J1001+0220, where environmental processing and quenching are further developed. We therefore interpret PC~J1001+0214 as a system in an active growth phase, with properties intermediate between highly extended, still-assembling overdense structures and more mature systems in which galaxy populations have been more strongly transformed by their environment.

Beyond the characterization of a single overdense structure, the discovery of PC~J1001+0214 has broader significance. It adds a valuable spectroscopically validated protocluster to the still limited sample of $z\sim3$ systems for which environmental signatures can be examined through direct comparison with a coeval field population ,providing an important example of a growth-dominated protocluster phase. The fact that such a system can still be identified in the COSMOS field, despite its extensive multiwavelength coverage and spectroscopic legacy, indicates that the census of high-redshift overdense structures remains incomplete even in the well-studied survey fields. This highlights the discovery space opened by combining the high-quality photometric redshifts of COSMOS2025 with the efficient spectroscopic validation enabled by COSMOS-3D, a strategy that can be extended to build a larger and more systematically characterized sample of protoclusters across different evolutionary stages.

For further research, investigating the chemical properties of galaxies in this structure is of value. In the current COSMOS-3D/NIRCam WFSS data, the spectroscopic identification mainly relies on He\,{\sc i}\,\(\lambda10830\), which is useful for redshift confirmation but does not provide robust gas-phase metallicity diagnostics. Future observations targeting rest-frame optical nebular lines such as H$\alpha$, H$\beta$, and [O\,{\sc iii}]$\lambda\lambda4959,5007$ will enable direct constraints on metallicity and ionization conditions, offering a decisive, quantitative test of how the protocluster environment regulates the baryon cycle and structural growth of its constituent galaxies at cosmic noon.
\citep{result_Z_Kewley_2002,results_Z_Maiolino_2019}.

\begin{acknowledgments}
This work is supported by National Key R\&D Program of China (grant no. 2023YFA1605600). This research is supported by National Natural Science Foundation of China (\#12525303) and Tsinghua University Initiative Scientific Research Program. This work is also funded by New Cornerstone Science Foundation through the XPLORER PRIZE.

This work is based in part on observations made with the NASA/ESA/CSA James Webb Space Telescope. The spectroscopic data were obtained from the Mikulski Archive for Space Telescopes at the Space Telescope Science Institute associated with program \#5893. The authors acknowledge the COSMOS2025 and COSMOS-3D teams for making this work possible through their efforts in building the photometric catalog and spectroscopic data products used in this study. 

\end{acknowledgments}

\facilities{JWST (NIRCam)}

\appendix
\section{SUPPLEMENTARY FIGURES}
\begin{figure}
    \centering
    \includegraphics[width=\linewidth]{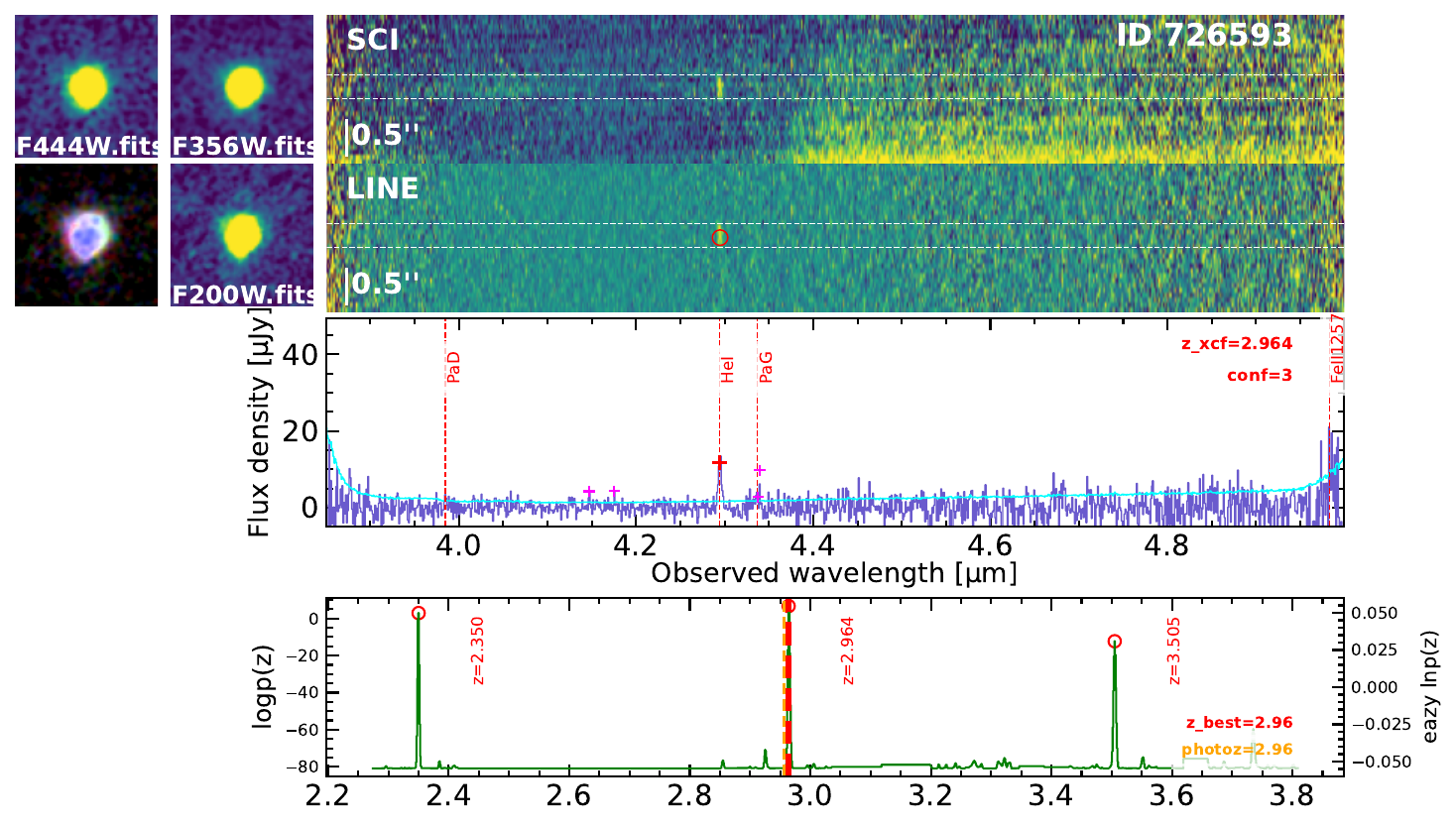}
    \caption{Example of a spectroscopically confirmed member in PC~J1001+0214 at $z_{\rm spec}=2.964$ (ID~726593). The left panels show image cutouts in F444W, F356W, a color-composite image, and F200W. The upper-right panels present the original two-dimensional grism spectrum and the continuum-subtracted line-only spectrum from the F444W grism. The white dashed lines mark the extraction aperture, and the red circle indicates the detected emission feature in the 2D line map. The middle panel shows the extracted one-dimensional spectrum, where the purple curve is the measured flux density and the cyan curve indicates the associated error. Red and magenta corsses mark detected emission features with ${\rm S/N}>5$ and ${\rm S/N}<5$, respectively, while the red dashed vertical lines indicate the expected positions of common emission lines at the best-fit grism redshift. The bottom panel shows the redshift-likelihood distribution from the cross-correlation analysis. The red dashed line marks the adopted spectroscopic redshift, and the orange dashed line indicate the photometric redshift. We refer to \citet{Lin_2026} for more details on the emission line searching algorithm.}
    \label{fig:grism}
\end{figure}

\begin{figure}[!htbp]
  \centering
  \begin{minipage}[t]{0.49\textwidth}
    \centering
    \includegraphics[width=\linewidth]{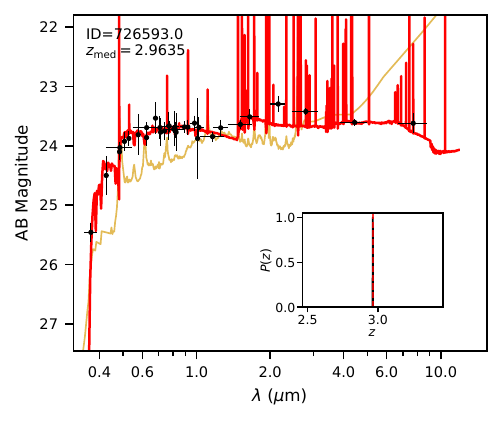}
  \end{minipage}\hfill
  \begin{minipage}[t]{0.49\textwidth}
    \centering
    \includegraphics[width=\linewidth]{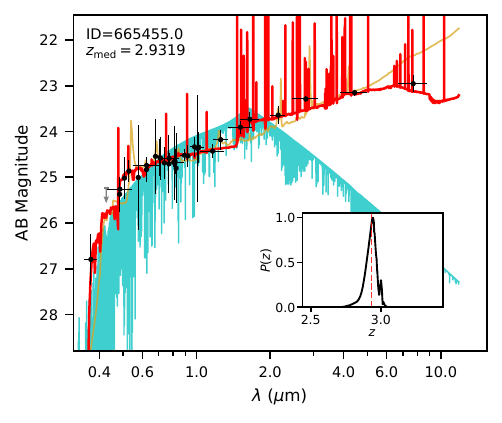}
  \end{minipage}

  \caption{Best-fit LEPHARE spectral energy distributions for two representative member galaxies in PC~J1001+0214. The left panel shows the spectroscopically confirmed member ID~726593, and the right panel shows the photometric member ID~665455. In each panel, the black points denote the observed photometry. The vertical errorbars show the 1$\sigma$ photometric uncertainties, and the horizontal bars indicate the wavelength ranges of the corresponding filters. The red curve shows the best-fit galaxy template, while the yellow and blue curves represent the AGN and star templates, respectively. The inset displays the Bayesian redshift probability distribution function, $P(z)$, and the dashed vertical line marks the adopted median photometric redshift, $z_{\rm med}$. We fix the $z_{\rm spec}$ derived from grism spectra for spectroscopic membes, while we let redshift vary in $2.5<z<3.5$ for photometric members.}
  \label{fig:lephare}
\end{figure}

\begin{figure}[!htbp]
  \centering
  \begin{minipage}[t]{0.49\textwidth}
    \centering
    \includegraphics[width=\linewidth]{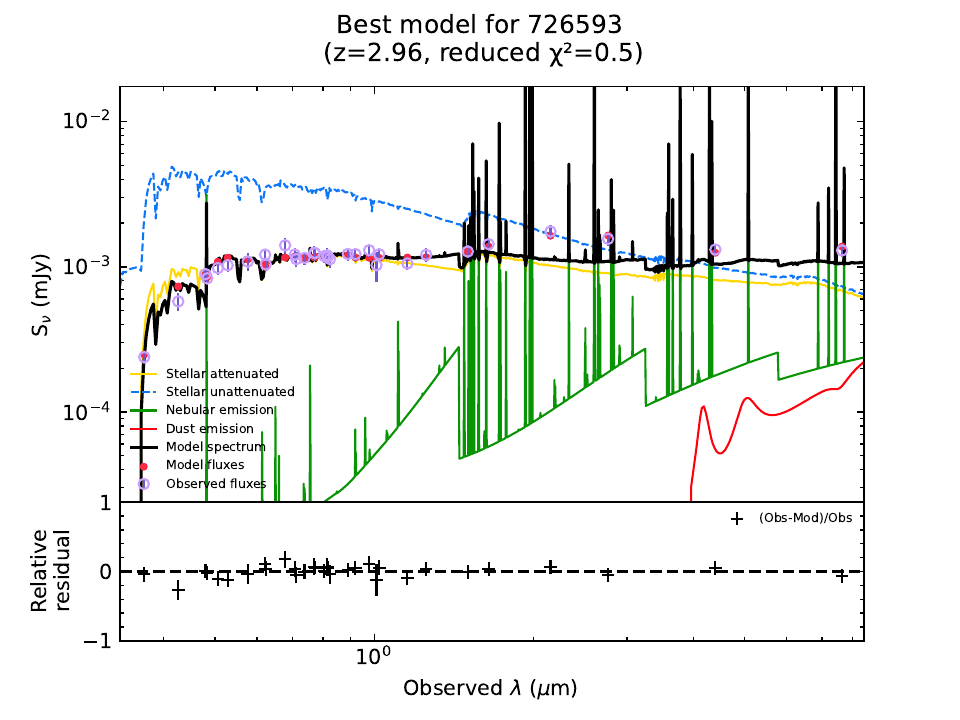}
  \end{minipage}\hfill
  \begin{minipage}[t]{0.49\textwidth}
    \centering
    \includegraphics[width=\linewidth]{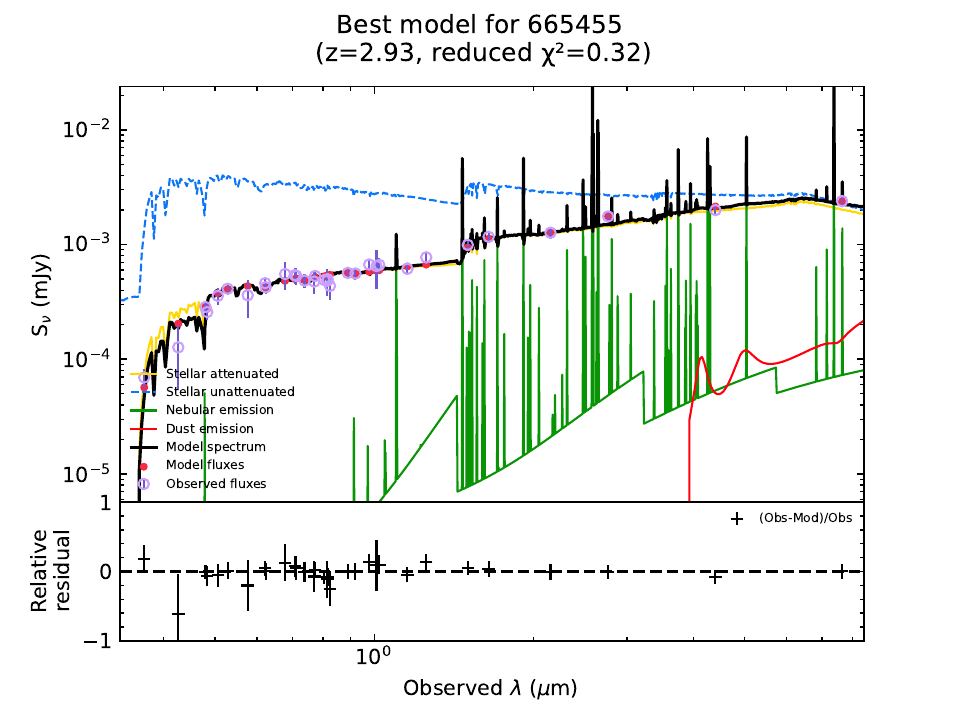}
  \end{minipage}

  \caption{Best-fit CIGALE spectral energy distributions for two representative member galaxies in PC~J1001+0214. The left panel shows the spectroscopically confirmed member ID~726593, and the right panel shows the photometric member ID~665455. In each panel, the open purple circles denote the observed fluxes, and the filled red circles indicate the model fluxes integrated through the corresponding filters. The black curve shows the total best-fit model spectrum. The yellow solid and blue dashed curves represent the attenuated and unattenuated stellar emission, respectively, while the green and red curves show the nebular and dust emission components. The lower panel in each plot presents the relative residuals, $(\mathrm{Obs}-\mathrm{Mod})/\mathrm{Obs}$. }
  \label{fig:cigale}
\end{figure}

Table~\ref{tab:matched_clusters} presents some previously reported protoclusters and high-redshift clusters matched to our overdensity candidates. Figure~\ref{fig:grism} shows the grism spectra and determined redshift of one spectroscopic member following the methods descirbed in Section~\ref{sec2.3} with its $z_{\rm spec}$ cross-matched against COSMOS compilation in \citet{compilation_Khostovan_2026}. Figure~\ref{fig:lephare} and Figure~\ref{fig:cigale} present the LEPHARE and CIGALE SED fitting results of one spectroscopic member (ID 726593) and one photometric member (ID 665455).

\begin{table}
\centering
\caption{Matched known protoclusters and high-redshift clusters in the literature for our candidates.}
\label{tab:matched_clusters}
\begin{tabular}{lccccc}
\hline\hline
Name & R.A. & decl.& $z$ & our $\delta$ & Reference \\
 & (deg.) & (deg.) &  &  &  \\
\hline
Hyperion-4      & 150.255600 & 2.342300 & 2.469 & 2.618 (0.58\%) & \citet{intro_Cucciati_2018} \\
Hyperion-3      & 149.999600 & 2.253700 & 2.444 & 2.513 (0.79\%) & \citet{intro_Cucciati_2018} \\
Hyperion-7      & 149.958100 & 2.218700 & 2.423 & 2.378 (1.24\%) & \citet{intro_Cucciati_2018} \\
CL J1001+0220   & 150.238042 & 2.336619 & 2.506 & 2.299 (1.54\%) & \citet{CL_J1001+0220_Wang_2016} \\
Hyperion-5      & 150.229300 & 2.338100 & 2.507 & 2.299 (1.54\%) & \citet{intro_Cucciati_2018} \\
ZFOURGE/ZFIRE   & 150.094000 & 2.251000 & 2.095 & 2.279 (1.62\%) & \citet{QG_Edward_2023} \\
LATIS2-D2-02    & 149.978000 & 1.970000 & 2.683 & 2.047 (2.71\%) & \citet{LATIS_Newman_2025} \\
LATIS2-D2-04    & 150.332000 & 2.274000 & 2.457 & 1.776 (4.68\%) & \citet{LATIS_Newman_2025} \\
Elent\'ari-P2     & 149.877400 & 2.285000 & 3.341 & 1.642 (6.01\%) & \citet{Elent_Forrest_2023} \\
Elent\'ari-P3     & 149.936900 & 2.274900 & 3.269 & 1.574 (6.70\%) & \citet{Elent_Forrest_2023} \\
Hyperion-6      & 150.331600 & 2.242700 & 2.492 & 1.562 (6.86\%) & \citet{intro_Cucciati_2018} \\
\hline
\end{tabular}

\tablecomments{
The table lists some previously reported protoclusters matched to our overdensity candidates.
Columns 2--4 give the literature coordinates and redshift of the matched structure from literature.
Column 5 gives the overdensity significance computed in this work for the corresponding candidate region, together with its rank among all $\delta$ values. The matched known structures are all associated with highly ranked overdensity peaks.
}
\end{table}

\section{Separate Comparisons of Spectroscopic and Photometric Members with the Field}
\begin{figure}[!htbp]
  \centering
  \begin{minipage}[t]{0.49\textwidth}
    \centering
    \includegraphics[width=\linewidth]{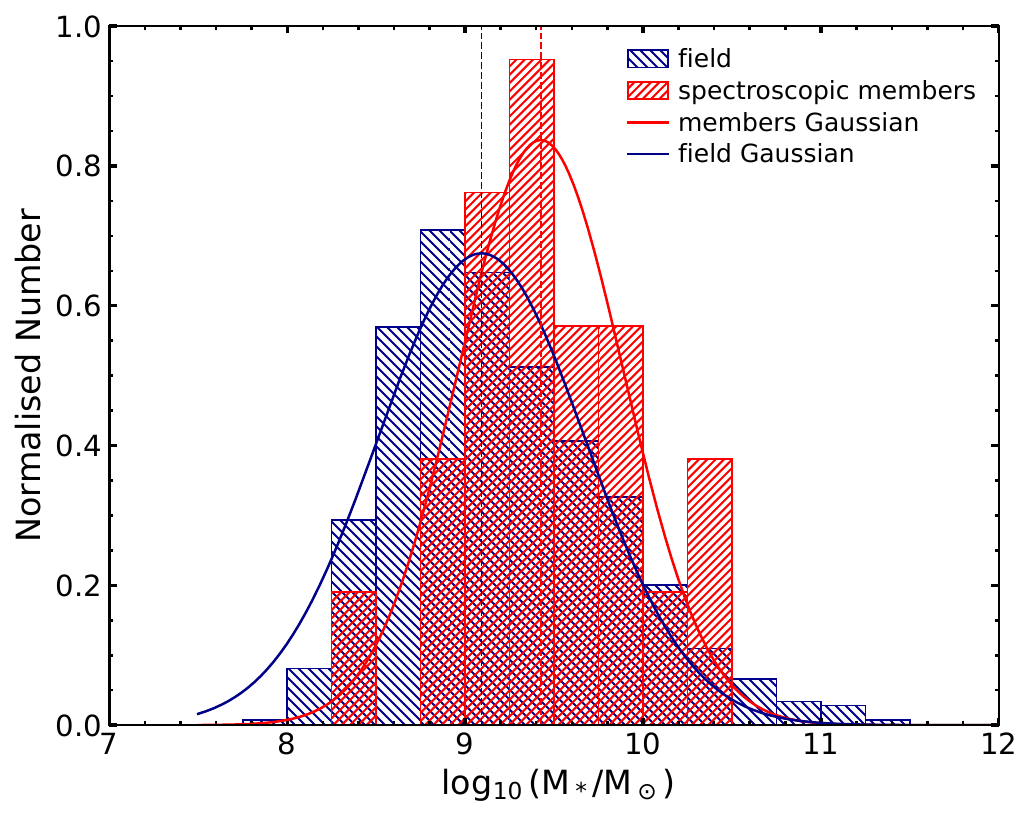}
  \end{minipage}\hfill
  \begin{minipage}[t]{0.49\textwidth}
    \centering
    \includegraphics[width=\linewidth]{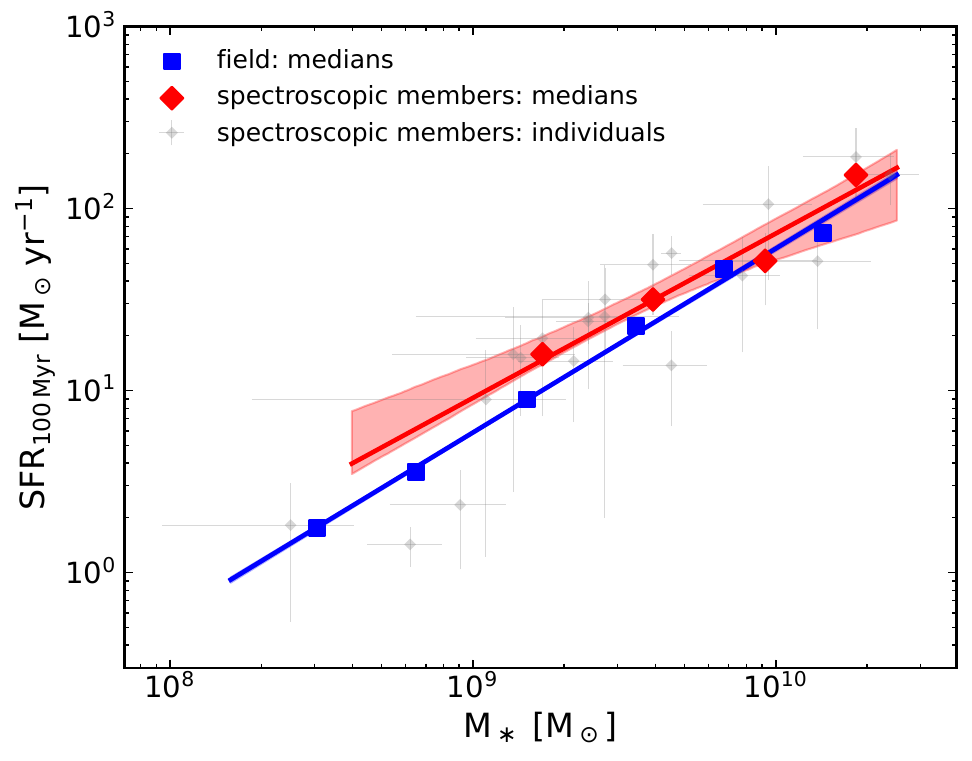}
  \end{minipage}

  \caption{Comparison between the spectroscopically confirmed members and the coeval field population. Left: normalized stellar mass distributions with Gaussian fits overplotted. Right: SFR$_{\rm 100\,Myr}$--$M_\star$ relation. The gray symbols show individual spectroscopic members with measurement uncertainties, while the blue squares and red diamonds denote the median median SFRs of the spectroscopic members and field sample in
different stellar mass bins, respectively. The corresponding curves show the best-fit relations for the two populations. The
corresponding error bars are estimated by bootstrap resampling.
 }
  \label{fig:spec_member}
\end{figure}

\begin{figure}[!htbp]
  \centering
  \begin{minipage}[t]{0.49\textwidth}
    \centering
    \includegraphics[width=\linewidth]{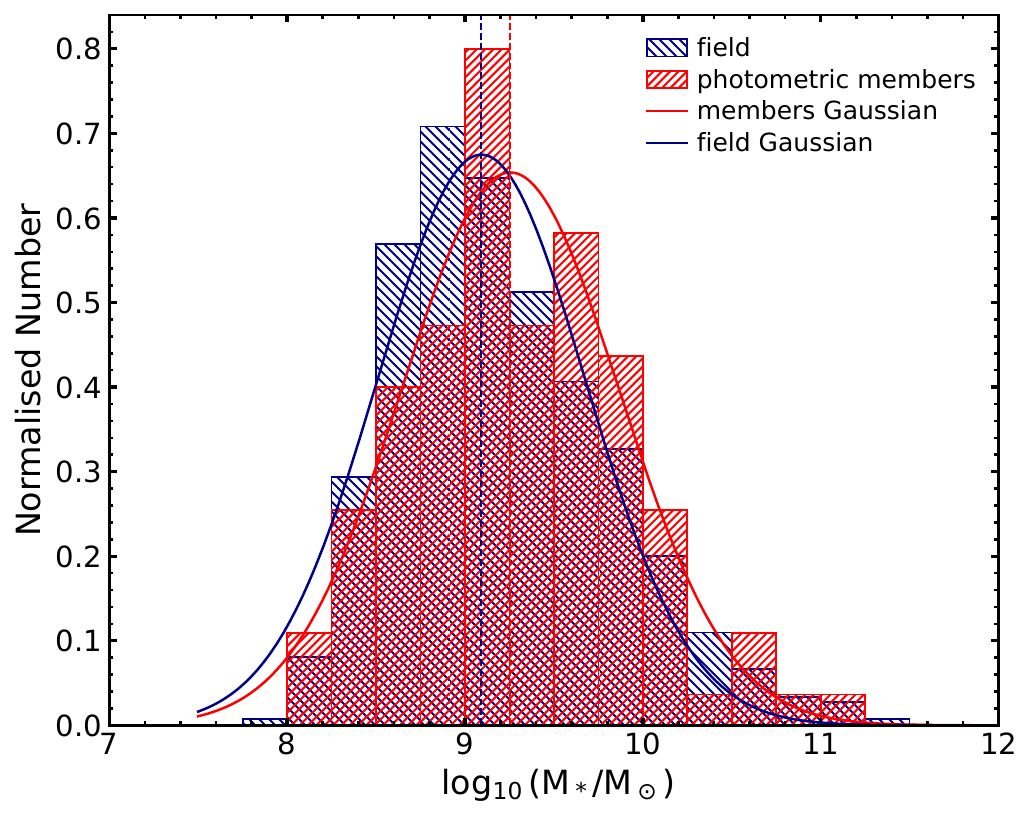}
  \end{minipage}\hfill
  \begin{minipage}[t]{0.49\textwidth}
    \centering
    \includegraphics[width=\linewidth]{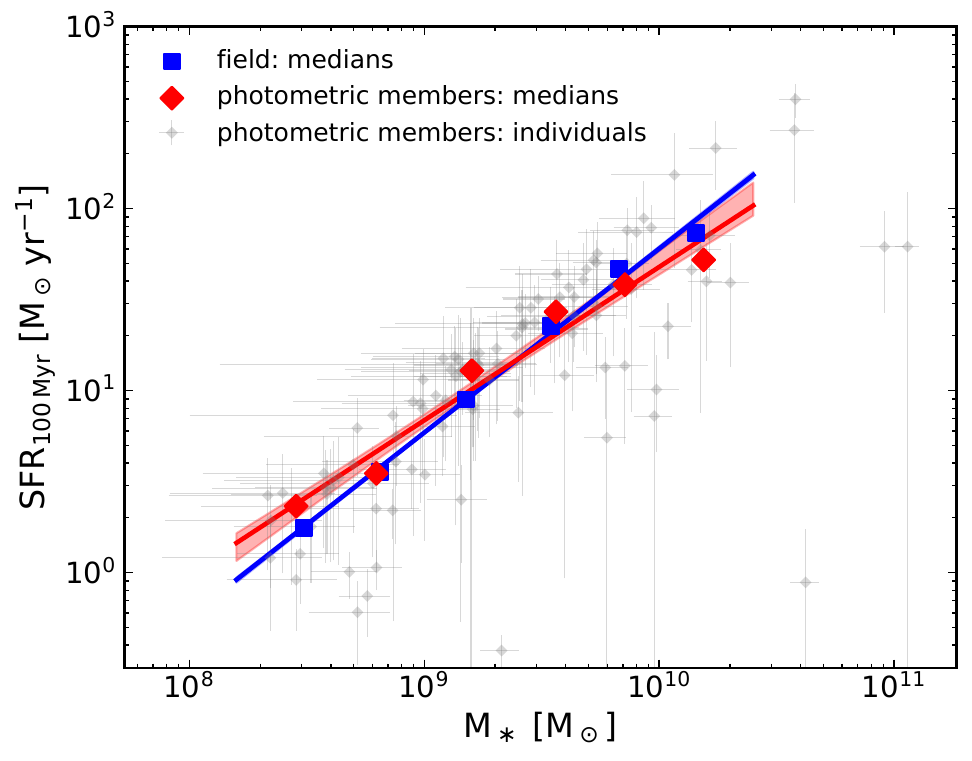}
  \end{minipage}

  \caption{Same as Figure~\ref{fig:spec_member}, but for the photometric member sample.
 }
  \label{fig:photo_member}
\end{figure}

In the main text, we analyze the protocluster member population by combining the 21 spectroscopically confirmed members with 110 photometric members. However, in our current work the spectroscopic identification mainly relies on He\,{\sc i}\,\(\lambda10830\).
The spectroscopic subsample may therefore be biased toward a physically distinct subset of the protocluster population and may not provide an unbiased representation of the full member population.
To assess this possibility, we compare the spectroscopic and photometric member subsamples separately with the coeval field population in Figure~\ref{fig:spec_member} and Figure~\ref{fig:photo_member} respectively.
We find that the two subsamples show broadly consistent behavior. Relative to the coeval field population, both the spectroscopic and photometric members tend to exhibit a stellar mass distribution shifted toward higher masses and an SFR--$M_\star$ relation elevated at the low-to-intermediate mass end. The trends reported in the main text are unlikely to be produced solely by this spectroscopically confirmed subset. Instead, they more likely reflect a general environmental signature of the protocluster population.

\section{Empirical Cumulative Stellar Mass Distribution}
\begin{figure}
    \centering
    \includegraphics[width=0.5\linewidth]{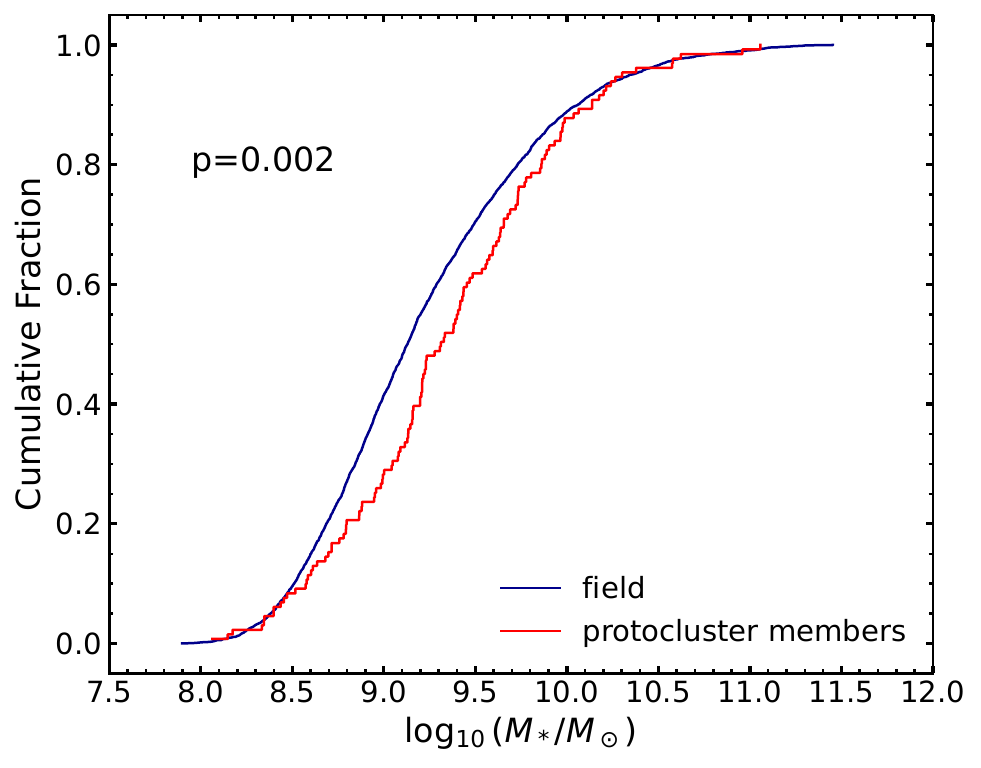}
    \caption{Empirical cumulative distribution functions (ECDFs) of stellar mass for the protocluster members and the coeval field galaxies. A two-sample Kolmogorov--Smirnov test gives $p=0.002$. }
    \label{fig:KS}
\end{figure}

To further quantify the difference between the stellar mass distributions of the protocluster members and the coeval field population, we compare their empirical cumulative distribution functions (ECDFs) in Figure~\ref{fig:KS}. The ECDF of the protocluster members is systematically shifted toward higher stellar masses relative to that of the field galaxies, consistent with the trend already suggested by the histogram-based comparison in the main text. We further perform a two-sample Kolmogorov--Smirnov (K--S) test on the two stellar mass distributions and obtain a p-value of $p=0.002$. This result indicates that protocluster population is skewed toward higher stellar masses than the coeval field population.

\bibliography{reference}{}
\bibliographystyle{aasjournalv7}

\end{document}